# A Computational Investigation of the Catalytic Properties of Graphene Oxide: Exploring Mechanisms Using DFT Methods

Danil W. Boukhvalov,[a],* Daniel R. Dreyer,[b] Christopher W. Bielawski[b] and Young-Woo Son[a]

Here we describe a computational study undertaken in an effort to elucidate the reaction mechanisms behind the experimentally observed oxidations and hydrations catalyzed by graphene oxide (GO). Using the oxidation of benzyl alcohol to benzaldehyde as a model reaction, density functional theory (DFT) calculations revealed that this reactivity stemmed from the transfer of hydrogen atoms from the organic molecule to the GO surface. In particular, neighbouring epoxide groups decorating GO's basal plane were ring-opened, resulting in the formation of diols, followed by dehydration. Consistent with the experimentally-observed dependence of this chemistry on molecular oxygen, our calculations revealed that the partially reduced catalyst was able to be recharged by molecular oxygen, allowing for catalyst turnover. Functional group-free carbon materials, such as graphite, were calculated to have substantially higher reaction barriers, indicating that the high chemical potential and rich functionality of GO are necessary for the observed reactivity.

## Introduction

Applications of transition metal-based catalysts for various oxidation and hydration processes are challenged for several fundamental reasons, including difficulties associated with their removal, high cost,s and limited natural stockpiles.[1] One approach to circumventing these ongoing challenges is through the use of metal-free catalysts as substitutes. While organocatalysts have been effective in this role,[2] carbon-based catalysts derived from graphite, graphene, and other similar materials may serve as useful alternatives. Though most carbon materials are unfunctionalized, reactive moieties may be introduced through reaction with appropriate reagents. Capitalizing upon this enhanced surface chemistry, recent experimental results have demonstrated the high catalytic activity of graphene oxide (GO)[3,4] as well other functionalized carbon materials, such as nitrogen doped graphene,[5] in various oxidation and hydration reactions. Beyond exploring novel fundamental reactivity, employing GO as a catalyst is attractive from a practical perspective owing to the abundance of natural carbon sources, as well as the catalyst's low density, extensive chemical functionalization, hydrophilicity, low cost and ease of preparation.[6] Moreover, the manipulation of GO's chemical composition is possible through empirical variation of the oxidants and reaction conditions employed in its preparation,[7] but the use of a computational model for these processes is necessary to guide the development of GO and other similar graphene-based catalysts for use in various chemical reactions.

GO is typically prepared by reacting graphite with strong oxidants (e.g., $KMnO_4$) under acidic conditions.[7] As a result, the graphite surface and edges become decorated with a wide range of functional groups, including alcohols, epoxides, and carboxylic acids. Moreover, these functional groups have been shown to be reactive toward a wide range of small molecule species, such as the selective oxidation of alcohols to aldehydes.[3,4] Concurrent with the oxidation of the small molecule species, experimental results have demonstrated that a significant reduction of the GO occurs after several catalytic cycles.[3] Increasing the temperature and volume of alcohol led to near total reduction of GO, affording graphite-like agglomerates that exhibited high conductivity (up to 4600 S m$^{-1}$) and a high carbon-to-oxygen ratio (up to 30 : 1). Differing methods used to synthesize GO lead to varying ratios of epoxy and hydroxyl groups, and depending upon the reaction conditions employed, the overall extent of coverage by oxygen ranges from approximately 50 to 80% of the carbon atoms forming GO's basal plane.[7] Incomplete functionalization of the graphite surface can be attributed to lattice distortions imparted by hydroxyls and other surface-bound functional groups.[9] Experimental and theoretical results have demonstrated that the GO reduction process is dominated by the removal of epoxy and hydroxyl groups.[9,10] With this structural model in hand and under support of both theoretical and experimental results, we sought to understand and explain GO's observed reactivity in the aforementioned oxidation and hydration reactions. The described changes of the oxidation level of graphene oxide during a catalytic reaction and the requirement of the re-oxidation of graphene for the further turnover permit us to propose catalytic mechanisms similar to the Mars-van Krevelen-type processes on solid surfaces.[11]

## Results and Discussion

### Modeling the catalytic process

The first step of our modeling was to examine the energetics of the oxidation of $PhCH_2OH$ to benzaldehyde (PhCHO) using GO (75% coverage of the carbon atoms) (Fig. 1a). We examined all possible intermediate states of this reaction and found that the migration of hydrogen atoms from the –$CH_2OH$ moiety of

[a] Prof. Dr. D. W. Boukhvalov, Prof. Dr. Young-Woo Son
School of Computational Sciences
Korea Institute for Advanced Studies
Seoul, 130-722, Korea
Fax: (+82)2-958-3820
E-mail: danil@kias.re.kr
[b] D. R. Dreyer, Prof. Dr. C. W. Bielawski
Department of Chemistry and Biochemistry
The University of Texas at Austin
1 University Station, A1590, Austin, TX 78712, USA

PhCH$_2$OH to an epoxy group on the GO surface to be much more energetically favorable than other pathways. The reaction, as described by our model, resulted in the formation of a diol on the GO surface, and we found the product to be energetically favored (see Fig. 1b-c). The formation of dangling bonds (principally present as unstable radicals) on PhCH$_2$OH or on GO was found to be energetically unfavorable making this reaction step endothermic. The second reaction step, as described above, was the migration of another hydrogen atom from PhCH$_2$OH onto one of the hydroxyl groups on the GO surface, resulting in the formation of an equivalent of water (Fig. 1c). This process is similar to the experimentally and theoretically discussed recombination of the hydrogen adatom on graphene and graphite surface with vanishing of the radical-like states.[12] Over the course of the reaction, one epoxy group was removed from GO, in agreement with experimentally observed reduction of GO by the oxidation of PhCH$_2$OH.[3,8]

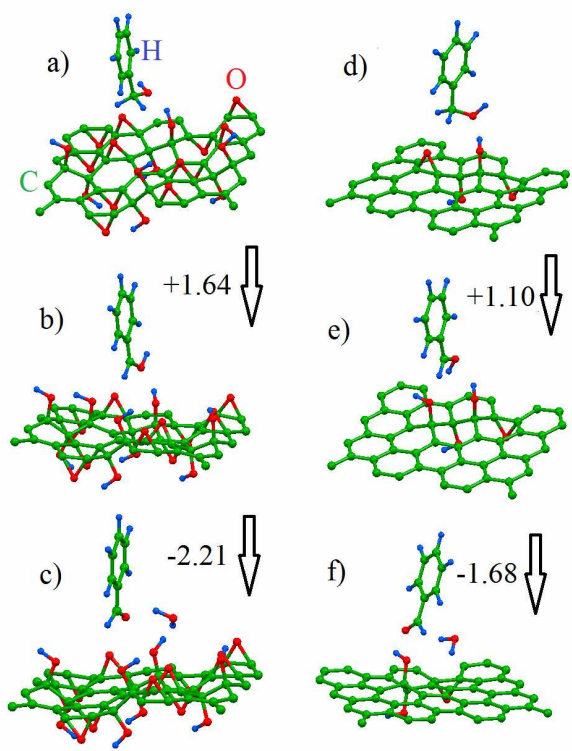

Figure 1. Optimized atomic structures for initial (a, d), intermediate (b, e) and final (c, f) steps of the oxidation of benzyl alcohol oxidation over GO with initial coverage of epoxy and hydroxyl groups 75% of the carbon atoms on the material's basal plan (a-c) and 12.5% (d-f). The total energy costs of reactions reported in eV.

With these preliminary mechanistic insights in hand, the next step of our modeling was to study the role of the level of reduction on the catalytic properties of GO. We performed the previously described calculations for the case of lower coverage (12.5 atom%) of GO by hydroxyl and epoxy groups (Fig. 1d–f), corresponding to a C : O ratio of 12 : 1. In these calculations, we found that the energy cost of the intermediate step of the reaction was of the same order as results described in the case of 75 atom% coverage. (Fig. 1e). Of note, however, the energy cost of the reactions was below 1.7 eV, indicating that the oxidation of PhCH$_2$OH could be realized in the presence of GO at 100 °C.

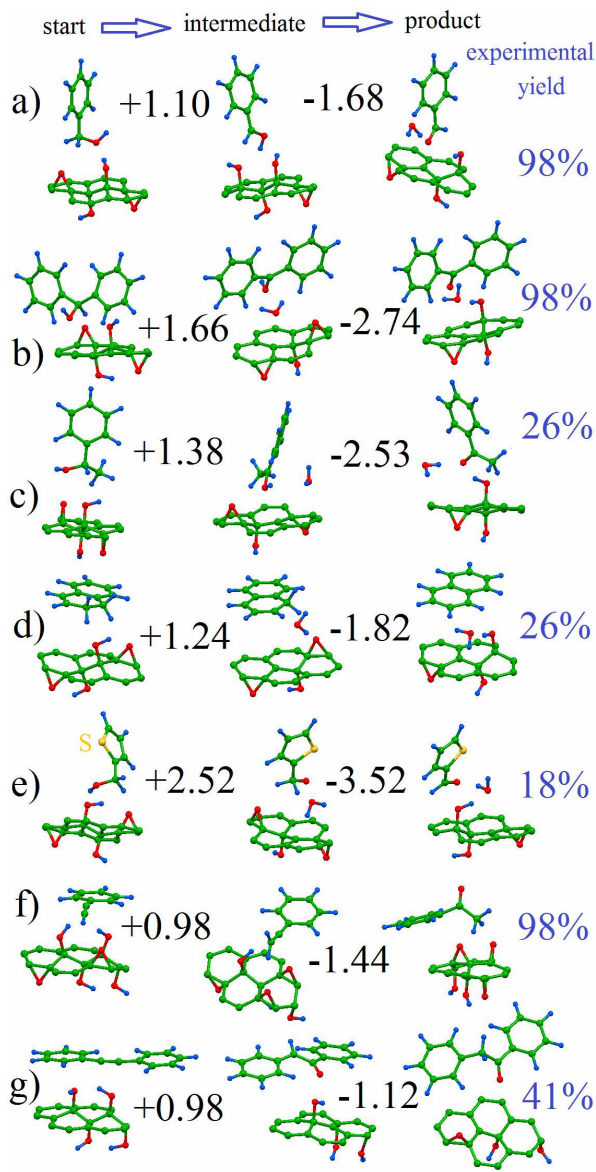

Figure 2. Optimized atomic structures of part of the GO lattice with low amount of hydroxyl and epoxy groups (see Fig. 1d) for the initial (left column), intermediate (central column) and final (right column) oxidation products (a: benzaldehyde from benzyl alcohol; b: benzophenone from diphenylmethanol; c: acetophenone from 1-phenylethanol; d: naphthalene from 1,2-dihydronaphthalene; e: 2-thiophenecarboxaldehyde from 2-thiophenemethanol) and hydration (f: acetophenone from phenylacetylene; g: phenyl benzyl ketone from diphenylacetylene) of various compound. The energy costs of the first and second steps of reactions are shown in eV.

The reduction of GO over the course of the reaction decreased the energy cost, but significantly diminished the number of active sites on the surface of the catalyst. For the case of unreduced GO (75 atom%) (Fig. 1a–c), six catalytic reactions could occur simultaneously over the model supercell, in contrast to two possible simultaneous reactions over reduced GO (Fig. 1d–f). Collectively, the mechanism described herein leads to the removal of epoxy groups from the surface of GO. In our 48 atom model, removal of all of these groups from the GO led to a C : O ratio of 24 : 1; very close to the experimental value (30 : 1) obtained by reduction of GO using alcohols.[8] These results led us to conclude that the reaction ceases when GO is free of surface epoxy groups.

To further explore the ability of the proposed model to probe the reactivity and mechanisms at work, we performed calculations of the total energies for other experimentally reported oxidation (Fig. 2b–f) and hydration (Fig. 2g–h) reactions performed using GO. In order to minimize the computation costs, we performed these calculations using partially reduced GO (12.5 atom% coverage). For three of the reactions (Fig. 2a–b,e), the low energy costs were in agreement with high reaction conversion. In the case of the oxidation of 2-thiophenemethanol to the corresponding aldehyde (Fig. 2e), the high energy cost correlated with the low experimental reaction conversion. In the case of the hydration of diphenylacetylene to benzyl phenyl ketone (Fig. 2g), the energy cost of the reaction was found to be nearly the same as the dehydration of benzyl phenyl ketone to diphenylacetylene over GO catalyst. This similarity in energies could explain the experimentally observed conversion of approximately 50%.[13]

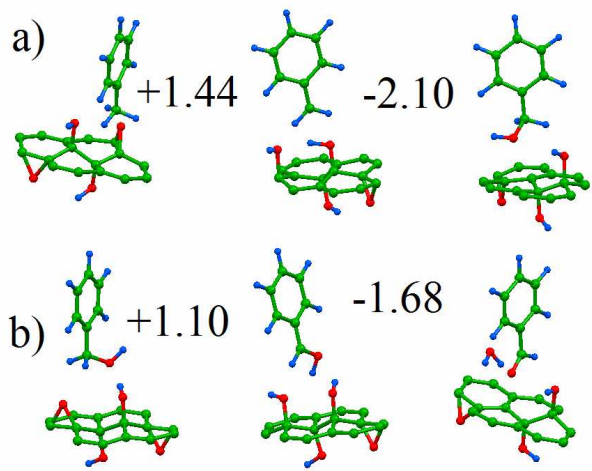

Figure 3. Optimized atomic structures of part of the GO lattice with low amount of hydroxyl and epoxy groups (see Fig. 1d) for first (a) and second (b) step of production of benzaldehyde from toluene via formation of benzene alcohol as intermediate product. The energy costs of the first and second steps of reactions are shown in eV.

The next step of our survey was to explore whether the proposed mechanisms could be applied to reactions consisting of multiple oxidation steps. We surmise that the production of PhCHO from toluene required the formation PhCH$_2$OH as an intermediate (see Fig. 3a), followed by further oxidation of this species to PhCHO (Fig. 3b). Such a step-wise oxidation has been reported for other oxygen-mediated transformations of toluene using both homogeneous and heterogeneous catalysts.[12] In our model, such a two-step reaction was found to require significant levels of GO oxidation; GO with 25 atom% coverage, or lower, was found to have insufficient epoxy groups for this reactions. The two-step oxidation also resulted in a two-fold faster reduction of the GO catalyst, localized to the vicinity of the reaction. The requirement for high catalyst oxidation and rapid deactivation of the catalyst explain both the relatively low energy cost of the reaction and the very low experimental yield (6%) obtained in some cases.[3]

**Regeneration of the graphene oxide catalyst**

In our earlier study of the oxidation of PhCH$_2$OH, we found the reaction to be dependent on the presence of an ambient atmosphere, suggesting that molecular oxygen was the terminal oxidant. We reasoned that ambient oxygen was in situ able to re-oxidize the GO as the catalyst was reduced during the reaction, as previously described, resulting in partial restoration of the active species. The last step of our modeling was to investigate this "recharging" of the GO catalyst in the presence of air and water. To this end, we calculated the energy cost of the chemisorption of additional oxygen from the air onto the GO surface. For this calculation, we added pairs of epoxy groups (formed through decomposition of molecular oxygen over the catalyst)[15,16] in several probable sites on one side of reduced GO (see Fig. 4a–b) and calculated the formation energy.

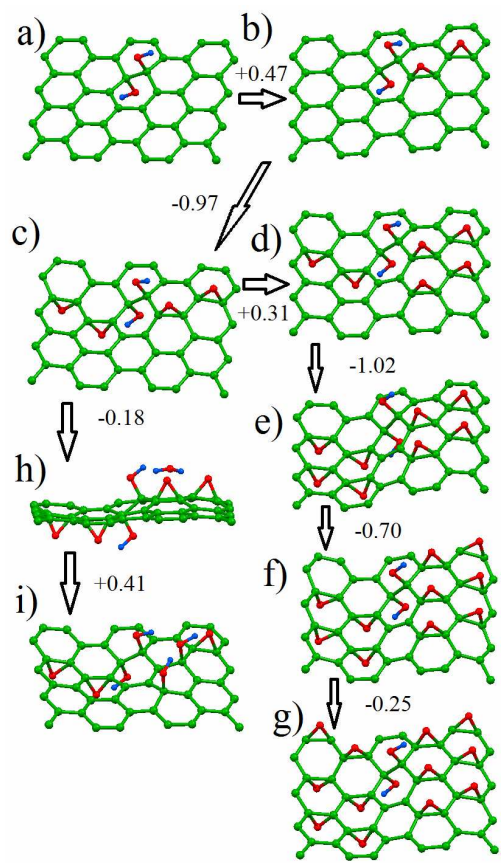

Figure 4. Optimized atomic structure of step-by-step of the oxidation of strongly reduced GO (a) with a C : O ratio 24 : 1 in the presence of molecular oxygen (b-g), where sp$^2$-hybridized C=C bonds are oxidized, resulting in the formation of new epoxide moieties. In the presence of both oxygen and water (c, h, i), these epoxide moieties can be hydrolyzed to form diols upon formation. The energy costs of each step, shown in eV, reveal that both are exothermic.

In addition to performing the aforementioned calculations, we also evaluated the energy[16] required for the activation of molecular oxygen on functionalized graphene and found that the presence of hydroxyl groups significantly decreased the activation barrier from 1.4 eV to 0.8 eV/O$_2$, similar to the case of N-doped graphene.[15] Further increasing the amount of epoxy groups to 50 atom% coverage decreased the oxygen activation energy to a minimum of 0.5 eV/O$_2$. Next, we added additional pairs of epoxy groups and calculated the formation energies for the more energetically favorable positions (Fig. 4c–g). For comparison, the calculated chemisorption energy for oxygen onto the reduced GO bearing hydroxyl groups was similar to the decrease in energy for the case of nitrogen doped graphene where the chemisorption energy was negative (in contrast to +1.35 eV for pure graphene).[17,18] Increasing the number of layers of the reduced GO led to a rise in oxygen chemisorption energy[18] and temperature of oxidation (from 200 °C for monolayer to 700 °C for 7-layer).[19] Thus in contrast to chemically inert graphite or pristine graphene, monolayer, oxidation of imperfect graphene was found to require very little energy.

The presence of hydroxyl groups in reduced GO allows for the formation of hydrogen bonds with water molecules (Fig. 4h). The formation of new hydroxyl groups from epoxy groups on the surface of GO, as a result of the mechanism previously described, required less than 0.5 eV (Fig. 4i). The values obtained for the oxidation and hydration processes of reduced GO were close to the energy required for water evaporation, defined as the difference between the formation energies of water in the liquid and gaseous phases (43.98 kJ/mol[20] or 0.46 eV). Thus, considering the energetics alone, these calculations showed that reduced GO may be re-oxidized at room temperature (or very rapidly re-oxidized at 100 °C) in the presence of water and air, as suggested by the experiment.[3]

## Conclusion

In summary, here we have described a series of computational studies focused on understanding the oxidation and hydration chemistry that may be performed using GO as a heterogeneous carbon-based catalyst. Employing DFT methods, we have found that the oxidation reactions (using the transformation of PhCH$_2$OH to PhCHO as a model) likely proceeded *via* transfer of hydrogen atoms from the organic starting material (PhCH$_2$OH) to the GO surface. In particular, the hydrogen atoms were believed to react with the epoxide moieties on the heavily functionalized GO surface, stemming from the known chemical instability of these species. Upon transfer of the hydrogen, the epoxides were converted into diols that were able to undergo further dehydration. The calculated energy costs of a range of organic molecules reveal the processes to be endothermic, and the energy requirements correlated well with the experimentally determined reaction conversions. Coupled with the oxidation of the organic species, the catalyst was found to undergo reduction during the reaction, resulting in an increase in the material's C : O ratio. This reduction process significantly decreased the yield of reactions comprising two or more oxidation steps (e.g., toluene to PhCH$_2$OH to PhCHO) due to a significant decrease in the number of catalytic sites in the vicinity of the reaction.

Consistent with our previous experimental observation that the oxidation reactions proceeded favourably in the presence of molecular oxygen, our calculations show that the catalyst may undergo re-oxidization under an atmosphere of air. Such a process allows for catalyst turnover and the completion of a catalytic cycle. The activity of GO in these oxidation and hydration transformations was also found to depend heavily upon the imperfections present on the catalyst's surface. The energy costs for the reactions were considerably higher when performed using graphite or pristine graphene in place of GO.

The computational results described herein, as well as our previous experimental results, indicate that GO possesses high chemical potential and a propensity to participate in chemical reactions. Importantly, these results have implications for a range applications of GO, such as energy storage devices, where GO's chemical reactivity could alter the host environment. Thus, these findings suggest that the continuing development of carbon materials may require divergent approaches: applications of carbon nanomaterials in areas that rely on chemical inertness and long term stability may find greater success when ultrahigh purity, functional group-free sources are utilized. In contrast, reaction catalysis and other areas that benefit from or require high chemical reactivity are likely to be advanced through the use of highly functionalized carbon materials, such as GO, rather than pristine forms such as flake graphite or graphene. Moreover, we anticipate that the precise mechanism proposed herein will help guide the development of more active and selective carbon-based catalyst through control of the catalyst's surface functionality (both identity and distribution).

## Computational models and methods

In order to examine the mechanisms of the catalytic oxidation and hydration processes, we performed calculations of the total energies of possible intermediate reaction steps. During these intermediate steps, hydrogen atoms from the small molecule species are proposed to migrate to the GO surface, transforming the surface-bound epoxy groups to alcohols (see Fig. 1a–b), which may then undergo further dehydration, or convert hydroxyl groups directly into water molecules (see Fig. 1b–c). Hydrogen atoms from the hydroxyl groups of GO may also migrate to the organic molecules. Such an abstraction would result in the formation of an oxy-centered radical which could then combine with an adjacent unsaturated carbon atom on the basal plane leading to the generation of epoxy groups on GO's surface. The calculated difference between the total energy of the system before and after the described processes is defined as the energy cost of the reaction. For the reactions presented in Fig. 1, we also performed calculations of the energetics of the transition steps between the initial (Fig. 1a, d) and intermediate steps (Fig. 1b, e), and intermediate and final (Fig. 1c, f) steps by the method described in detail in previous works.[19] Similar to the migration of hydroxyl groups and hydrogen atoms across the graphene surface, a greater part (about 0.9 eV) of the energy required for migration of hydrogen from benzyl alcohol to GO corresponded to the first transition step of migration. Breaking the bond between carbon or oxygen and the hydrogen, and further migration of the atom required less that 0.2 eV of energy. Because the obtained energies of the transition steps were lower than the energies required for the formation of the intermediate step (Fig. 1b), we will not discuss further the energetics of the other intermediate steps.

In contrast with previous work,[22] we do not consider the role of edges in our present model for three primary reasons. First, edge-bound oxygen groups have been shown both theoretically[7] and experimentally[21] to be exceptionally stable and chemically inert. In contrast, oxygen groups bound to the surface of GO's basal plan are far more labile, resulting in group migration or desorption, making these species significantly more reactive toward chemical species. Second, we found there was high similarity between the activation energy required for migration of epoxy groups and the energies of oxygen desorption from graphene. The latter of these processes would lead to removal of oxygen functional groups from GO as O$_2$ molecules, instead of migration of these groups to GO's edges.[9,14,15,20] Finally, the high aspect ratio of GO results in an insignificant number of active sites on the edges, as compared to the much larger basal plane of the GO sheets. For example, exhaustive reduction of GO using benzyl alcohol (PhCH$_2$OH) as a reductant led to a final C : O ratio of 30 : 1 .[8] If we consider only the edge groups, this ratio corresponds to a graphene sheet of approximately 12 nm, but the lamellae of GO are typically much larger (hundreds of nm to microns, depending on the particle size of the starting flake graphite source).[7] Thus, oxygen groups not only at the edges but also in the central part of GO sheets are likely participate in the reduction process, and by extension, the catalytic reactions.

In the present study, modeling was performed using density functional theory (DFT) in the pseudopotential code SIESTA,[23] as was done in our previous work.[9,17] All calculations were performed using the generalized gradient approximation (GGA-PBE[24]) which is feasible for the description of strong covalent bonds.[18] We omit the underestimation of the weak van der Waals bonds within GGA because the values of the binding energies these types of bonds are typically below 0.1 eV. Full optimization of the atomic positions was performed. During the optimization, the ion cores were described by norm-conserving pseudo-potentials.[27] The wavefunctions were expanded with a double-ζ plus polarization basis of localized orbitals for carbon and oxygen, and a double-ζ basis for hydrogen. Optimization of the force and total energy was performed with an accuracy of 0.04 eV/Å and 1 meV, respectively. All calculations were carried out with an energy mesh cut-off of 360 Ry and a k-point mesh of 8 × 8 × 1 in the Mokhorst-Park scheme.[26] For our modeling, we have used a rectangular supercell containing 48 carbon atoms and two hydroxyl and two epoxy groups on its surface.[9] The spatial distribution of these groups was based on our previous modeling and corresponded to the maximal level of coverage by hydroxyl and epoxy groups.[9] In order to estimate the connection between the calculated energy cost of the reactions and the probability of the process occurring at experimental temperatures (100-200 °C), we compared the calculated energy required for the oxygen chemisorption and activation on the monolayer graphene (about 1.4 eV)[16-18] with the experimental results of graphene monolayer fast, large area oxidation below 200 °C. [19] The calculation of formation energies for the re-oxidation of reduced GO was performed using Eq.1:

$E_{form} = E_{GO} - [E_{rGO} + E_{O2}]$                Eq. 1

where $E_{GO}$ is the total energy of the re-oxidized GO, $E_{rGO}$ is the total energy of reduced GO, and $E_{O2}$ is the total energy of the oxygen molecule in empty box. The interaction of GO with water during re-oxidation can be described by a similar formula.

## Acknowledgements

We acknowledge computational support from the CAC of KIAS. Y.-W. S. was supported by the NRF grant funded by MEST (Quantum Metamaterials Research Center, R11-2008-053-01002-0 and Nano R&D program 2008-03670). C.W.B. acknowledges the Welch Foundation for support (F-1621).

**Keywords:** graphene • graphene oxide • DFT • oxidation • hydration • carbocatalysis